# All-polarization-maintaining, single-port Er:fiber comb for high-stability comparison of optical lattice clocks

Noriaki Ohmae[1,2,3]*, Naoya Kuse[4], Martin E. Fermann[5], and Hidetoshi Katori[1,2,3,6]

[1]*Quantum Metrology Laboratory, RIKEN, Wako, Saitama, 351-0198, Japan*

[2]*Department of Applied Physics, Graduate School of Engineering, The University of Tokyo, Bunkyo, Tokyo, 113-8656, Japan*

[3]*Innovative Space-Time Project, ERATO, Japan Science and Technology Agency, Bunkyo, Tokyo, 113-8656, Japan*

[4]*IMRA America Inc., Boulder Research Laboratory, Longmont, CO, 80501, USA*

[5]*IMRA America Inc., Ann Arbor, MI, 48105, USA*

[6]*RIKEN Center for Advanced Photonics, Wako, Saitama, 351-0198, Japan*

E-mail: ohmae@riken.jp

All-polarization-maintaining, single-port Er:fiber combs offer long-term robust operation as well as high stability. We have built two such combs and evaluated the transfer noise for linking optical clocks. A uniformly broadened spectrum over 135-285 THz with a high signal-to-noise ratio enables the optical frequency measurement of the subharmonics of strontium, ytterbium, and mercury optical lattice clocks with the fractional frequency-noise power spectral density of $(1-2)\times10^{-17}$ Hz$^{-1/2}$ at 1 Hz. By applying a synchronous clock comparison, the comb enables clock ratio measurements with $10^{-17}$ instability at 1 s, which is one order of magnitude smaller than the best instability of the frequency ratio of optical lattice clocks.

Optical frequency combs have become indispensable tools for precision measurements including atomic/molecular spectroscopy, low-phase-noise microwave generation, and ranging.[1] In optical clocks, linking different atomic clocks[2,3] or distant clocks via optical fibers[4] with a fractional uncertainty of $10^{-18}$ is of significant concern with a future redefinition of the second in the International System of Units (SI)[5]. Such endeavors, in turn, offer intriguing opportunities for testing the constancy of the fundamental constants[6] and for relativistic geodesy[4,7]. These applications necessitate ultralow-noise optical frequency combs that allow long-term and robust operation. Titanium-sapphire-based frequency combs have demonstrated outstanding stability.[8,9] However, their bulky optical setups require regular maintenance, thus hampering long-term and robust operation, which limits their potential applications. In contrast, erbium (Er) fiber combs enable all-fiber architecture for robust operation. Although a typical Er:fiber comb uses nonlinear polarization rotation (NPR) to acquire mode-locking with excellent noise performance[9-13], the operational condition for such NPR-based Er:fiber oscillators can be sensitive to environmental conditions. For practical applications, such as the long-term operation of clocks to generate the optical second[14], and field and space applications, the all-polarization-maintaining (PM) architecture is preferred[15,16]. However, such architecture has shown relatively large phase noise. Low intrinsic phase noise with PM architecture is demonstrated by applying a nonlinear amplifying loop mirror (NALM)[17,18].

In linking multiple optical frequencies, Er:fiber combs with a multibranch configuration, where each port consists of an Er-doped fiber amplifier (EDFA) and a highly nonlinear fiber (HNLF), have been employed[3,19], as it allows sufficient output power per comb tooth optimized for the single frequency. In such a multi-branch comb, the phase noise in different branches introduces the instability of $\sim 10^{-16}$ at 1 s[10,11]. A record high instability of $4 \times 10^{-16} (\tau/\text{s})^{-1/2}$ has been demonstrated for synchronous clock comparison between strontium (Sr) and ytterbium (Yb)-based optical lattice clocks[3], in which the instability is mainly limited by the Dick effect[20] due to the frequency noise of the multibranch comb.

The single-port architecture[12,13] is advantageous for suppressing such interbranch relative phase noise that is caused by the optical path length fluctuation. Moreover, in order to access multiple optical clocks with different frequencies, an octave-spanning super-continuum (SC) output with a sufficient signal-to-noise ratio (SNR) is favored. In this work, we develop a low-noise and single-port Er:fiber comb by utilizing an NALM-based all-PM architecture. A uniformly broadened high-SNR comb over 135-285 THz allows linking Sr-, Yb-, and mercury (Hg)-based optical lattice clocks, which operate at 429-1129 THz, with a modified Allan deviation below $10^{-17}$ at $\tau = 1$ s. By applying synchronous operation, we show that an optical lattice clock comparison with $2\times10^{-17}(\tau/\text{s})^{-1/2}$ is possible, which is one order of magnitude smaller than the best instability of the frequency ratio of optical lattice clocks.[3]

Figure 1 (a) shows the schematic of the all-PM Er:fiber comb. Mode-locking based on NALM is self-starting. The Er:fiber oscillator (the lower box in the left) is equipped with an electro-optic modulator (EOM) and a piezoelectric transducer (PZT) glued along the fiber, both of which are used to control the repetition rate $f_{REP} \approx 80$ MHz. The oscillator has an average power of 1 mW with a spectral bandwidth of about 10 THz and the center frequency of 192 THz. Figure 2 shows the optical spectra of the Er:fiber oscillator and the SC output. This oscillator has a single-sideband phase-noise floor of about -100 dBc/Hz, which is measured by beating with a kHz-linewidth laser at 192 THz. Applying an EDFA and a HNLF, we obtain a uniformly broadened octave-spanning spectrum (red line in Fig. 2) with the average power of about 80 mW. To evaluate the SNR of this SC output, we simultaneously monitor the beat signals with lasers at 215, 259, and 282 THz, which respectively correspond to the subharmonics of Sr, Yb, and Hg clock frequencies[21]. Figure 1(b) shows the beat signals, all of which have the SNRs of larger than 30 dB measured with the resolution bandwidth (RBW) of 100 kHz or the phase-noise floors smaller than -80 dBc/Hz. This sets the noise floor of the frequency ratio measurement with the power spectrum density (PSD) of $1\times10^{-18}(f/\text{Hz})$ 1/Hz$^{1/2}$, where $f$ is the Fourier frequency of the noise component. The SNR degradation from the Er:fiber oscillator output to the SC output was observed to be less than 20 dB, which is in agreement with the spectral broadening of about 15 (=150 THz/10 THz). The carrier-envelop offset frequency $f_{CEO}$ is obtained using a self-referencing interferometer with the phase-noise floor of -100 dBc/Hz, as shown in the lower panel in Fig. 1(b).

The spectral shape and bandwidth of the SC output is optimized by the current supplied to pumping laser diodes (LDs) for the EDFA. Once optimized, the beat signals, as shown in Fig. 1(b), maintain their SNRs without daily adjustment, which demonstrates the long-term stable operation of the system. In about 85 % of the spectral region in the range of 135-285

THz, we observed a tooth power of > 10 nW per mode. The shot noise for the detected averaging power of 100 μW gives a phase noise floor of < -100 dBc/Hz. Electric noises caused in the photo detectors (PDs) and amplifiers typically give a phase noise floor of -90 dBc/Hz. Therefore, the measured phase noise floor of about -80 dBc/Hz is solely limited by the noise floor of the comb spectrum.

As both $f_{CEO}$ and the beat signals in free-running operation show intrinsic linewidths of < 0.1 MHz, feedback control with a unity-gain frequency of a few hundreds of kHz is sufficient for the tight locking of both $f_{CEO}$ and $f_{REP}$. The root-mean-square (RMS) value of the contributed phase noise (integrated phase noise to 5 MHz) calculated from a white phase noise floor of -80 dBc/Hz is $\left( \int_0^{5\,\text{MHz}} S_\phi(f) df \right)^{1/2} = 0.1 \ll \pi$ rad, where $S_\phi(f)$ is the phase noise PSD in units of rad$^2$/Hz. This small RMS phase noise allows the tight locking of $f_{CEO}$ and $f_{REP}$ to the references without the need for extra devices such as transfer oscillators to reduce the noise bandwidth or frequency dividers to expand the frequency capture range. In our experiment, we stabilize $f_{CEO}$ by controlling the current of the pump LD with a unity-gain frequency of 0.2 MHz, and $f_{REP}$ by using a PZT for a slow signal (DC – 1 kHz) and an EOM with a unity-gain frequency of 1.3 MHz. The residual contributed phase noise (integrated from 3 Hz to 5 MHz) of the in-loop beat signals for both $f_{CEO}$ and the beat signal at 215 THz is ≤ 0.2 rad, which is comparable to the values reported for non-PM NPR-based Er:fiber combs[10]. This residual phase noise is sufficiently small to remain locked for longer than a few days[17].

To evaluate the frequency spectral transfer noise, we prepare two identical combs, both of which are stabilized to the 'clock laser' at 215 THz, which is locked to a 40-cm-long reference cavity[21]. Lasers at 259 and 282 THz are stabilized to the respective teeth of Er comb (1) in Fig. 1(a). In the following, we analyze the beat signals of these lasers and Er comb (2) with the measurement bandwidth of 2 MHz by using in- and quadrature-phase demodulators based on analog frequency mixers and the RF reference. Most of the optical paths connecting the two combs with PM fibers and free-space optics are stabilized using interferometer-based Doppler noise cancellers (DNCs)[22], where an approximately 10-cm-long optical path remains uncompensated. Since the frequency noise of the spectral transfer via the comb typically shows white phase noise characteristics, we use the modified Allan deviation to indicate the instability. Figure 3 shows the modified Allan deviation of the fractional frequency noise observed in the spectral transfer from 215 to 259 THz (blue) and to 282 THz (red). This shows instabilities of $(5-7)\times10^{-18}$ at $\tau = 1$ s and $1\times10^{-19}$ at $\tau = 100$ s, which is similar to those described in Refs. 12 and 13. Compared with the instability using a multi-branch NPR-based Er:fiber comb[10,21] as shown by a black line, the spectral transfer instability at $\tau = 1$ s is improved by more than 30 times. At $\tau > 100$ s, the instability reaches the floor of around $10^{-19}$, which is most likely caused by the fluctuation of the residual uncompensated optical path length. Since this measurement utilizes an RF reference at 10 MHz with a fractional frequency uncertainty of $10^{-12}$, this corresponds to a sub-$10^{-19}$ uncertainty. Within this statistical uncertainty, there is no obvious frequency offset.

Figure 4(a) shows the fractional frequency noise PSD of the frequency spectral transfer. For $f > 100$ Hz, the spectral transfer noise is close to the limit estimated by the SNR of the beat signals except for a bump at around 1 kHz, which is caused by the insufficient control gain of DNCs. On the other hand, excess noise above the SNR limit is observed for $f < 100$ Hz. This is possibly due to the fluctuation of the uncompensated optical path length. Consequently, we achieved the frequency noise PSD of $(1-2)\times10^{-17}$ $1/Hz^{1/2}$ at 1 Hz for the spectral transfer between 215, 259, and 282 THz.

Finally, we discuss the fractional instability of the frequency ratio $R = \nu_1/\nu_2$ in clock comparison, which offers a benchmark for testing the short-term stability of the frequency comb used in atomic clocks. The fractional instability is limited by the quantum projection noise (QPN)[23] $\sigma_{\text{QPN}}$, the Dick effect $\sigma_{\text{Dick}}$[20], and the spectral transfer noise of the comb $\sigma_{\text{Comb}}$ to bridge the two clock frequencies $\nu_1$ and $\nu_2$. The overall instability is described as

$$\sigma_y \approx \left(\sigma_{\text{QPN}}^2 + \sigma_{\text{Dick}}^2 + \sigma_{\text{Comb}}^2\right)^{1/2}. \qquad (1)$$

The QPN-limited instability for respective clocks $j$=1,2 is given by $\sigma_{\text{QPN}j} \sim (\nu_j T_i)^{-1} \left(N_j \tau / T_C\right)^{-1/2}$, where $\nu_j$ is the transition frequency, $T_i$ the interrogation time of the clock transition, and $N_j$ the number of atoms interrogated in cycle time $T_C = T_i + 1$ s including the atom preparation time of 1 s, resulting in the total instability of $(\sigma_{\text{QPN1}}^2 + \sigma_{\text{QPN2}}^2)^{1/2}$ for two clocks. In optical lattice clocks interrogating $N_j > 10^3$ atoms, a QPN-limited instability better than $\sim10^{-16}(\tau/s)^{-1/2}$ is achievable, which is comparable to or smaller than the Dick effect limited instabilities[24]. Moreover, $\sigma_{\text{QPN}}$ can be further reduced by increasing $N_j$ and/or $T_i/T_C$.

The Dick-effect-limited instability $\sigma_{\text{Dick}}$ is caused by the down-conversion of the frequency noise of the 'clock laser' that periodically interrogates the clock transition with a deadtime,

$$\sigma_{\text{Dick}} = \left\{ \frac{1}{(\tau/s)} \sum_{n=1}^{\infty} \left[ \left( \frac{g_s^n}{g_0} \right)^2 + \left( \frac{g_c^n}{g_0} \right)^2 \right] S_y(n/T_C) \right\}^{1/2}, \qquad (2)$$

where $S_y(f)$ is the fractional frequency noise PSD in units of 1/Hz , $g_0$ is the 1-cycle average of a sensitivity function $g(t)$, and $g_c^n$ and $g_s^n$ are the cosine and sine components of the $n$-th Fourier series expansion of $g(t)$, respectively[20]. As the sensitivity $[(g_c^n/g_0)^2 + (g_s^n/g_0)^2]^{1/2}$ rapidly decreases for $f \gg 1/T_i$, frequency noise with low Fourier components of a few Hz solely affects the clock instability $\sigma_{\text{Dick}}$. Note that the frequency noise of the clock laser with the state-of-the-art instability $\sigma_y \sim 1 \times 10^{-16}$ at $\tau = 1$ s [25,26,27] [green line in Fig. 4(a)] is an order of magnitude larger than the frequency transfer noise of the comb for $f < 100$ Hz, where the Dick effect plays a decisive role. Assuming this laser noise, the Dick effect limit of comparison of Yb (Hg) and Sr optical lattice clocks is calculated to be $\sigma_{\text{Dick}j} \sim 10^{-16} (\tau/s)^{-1/2}$ for each clock $j$=1 and 2, as shown by the black circles in Fig. 4(b). This indicates that the thermal noise of an optical cavity[28] used to stabilize the clock laser severely degrades the short-term instability of optical clocks, and the superb transfer instability of the comb is not fully utilized.

The ratio measurement beyond the Dick effect limit of the 'clock laser' is possible by applying synchronous interrogation to reject the laser frequency noise[3,24]. When the two clocks share a single cavity by transferring its spectral characteristics via the comb, the Dick effect term $\sigma_{\text{Dick}}$ in Eq. (1) is given by

$$\sigma_{\text{Dick}}^{\text{Sync}} = \left( \sigma_{\text{Dick(Cavity)}}^2 + \sigma_{\text{Dick(Comb)}}^2 \right)^{1/2}, \qquad (3)$$

where the Dick effect due to the cavity-induced laser noise is partially rejected and reduced to $\sigma_{\text{Dick(Cavity)}}$, and the spectral transfer via the comb adds an extra Dick effect $\sigma_{\text{Dick(Comb)}}$. Employing a comb with lower frequency noise than the cavity thermal noise will allow $\sigma_{\text{Dick}}^{\text{Sync}} < \sigma_{\text{Dick}}^{\text{Async}}$, as demonstrated in Ref. 3, where the synchronous frequency ratio measurement of Yb and Sr optical lattice clocks is mainly limited by the Dick limit $\sigma_{\text{Dick(comb)}}$ of the multi-branch Er:fiber comb.

The green circles in Fig. 4(b) indicate the instability for synchronous comparison with our new Er:fiber comb. The cavity-related Dick effect $\sigma_{\text{Dick(Cavity)}}$ is more than two orders of magnitude (red circles) smaller than $\sigma_{\text{Dick}}$ and the comb frequency noise sets the Dick effect limit $\sigma_{\text{Dick(comb)}}$, allowing total instability of low $10^{-17}(\tau/\text{s})^{-1/2}$, which improves the stability by an order of magnitude compared with asynchronous operation (black circles). We assume $N_j = 10^5$ atoms to allow $\sigma_{\text{QPN}} \leq \sigma_{\text{Dick}}^{\text{Sync}}$ (orange line), which will be affordable with optical lattice clocks. The application of a cryogenic monocrystalline silicon cavity[27] or crystalline-coated mirrors[29] with reduced thermal noise of frequency noise PSD of ~3×10$^{-17}$($f$/Hz)$^{-1/2}$ 1/Hz$^{1/2}$ will allow further reduction of the instability. As for the last term in Eq. (1), $\sigma_{\text{Comb}}$ decreases faster than $\tau^{-1/2}$ for longer averaging time down to $10^{-19}$ [8,9,12], which is 10-100 times smaller than that of current clock comparisons.

In summary, we have developed all-PM and single-port Er:fiber combs with an octave-spanning high-SNR optical spectrum, which is, in particular, suitable for the high-stability ratio measurement of optical lattice clocks. By synchronously operating the clocks, we show that the Dick-effect contribution of the comb instability to low $10^{-17}(\tau/\text{s})^{-1/2}$ is achievable, which is even better than the state-of-the-art laser instability. Low-noise and all-PM architecture facilitates robust comparison of highly stable optical lattice clocks[4], offering new applications in cm-level[7] relativistic geodesy and a search for the variation of fundamental constants[30] and the Lorentz invariance[31] at shorter time scales.

## Acknowledgments

This work is partially supported by the Photon Frontier Network Program of the Ministry of Education, Culture, Sports, Science and Technology, Japan. We thank Dr. Takamoto and Dr. Nemitz for providing lasers of Sr and Yb clocks.

**Figure Captions**

**Fig. 1.** (a) Optical configuration of a single-port, all-PM Er:fiber comb and signal detections for frequency ratio measurement of the optical lattice clocks consisting of Sr, Yb, and Hg. Periodically poled lithium niobate (PPLN) is used for the self-referencing $f$-$2f$ interferometer. PR, partial reflectors; SHG, second-harmonic generator; BS, beam splitter; PD, photo detector; DBM, double-balanced mixer; OSC, oscillator. (b) RF spectra of beat signals at 215, 259, and 282 THz, and $f_{CEO}$ signal measured with RBW = 100 kHz in free-running operation.

**Fig. 2.** Optical spectra of the oscillator (blue) and the super-continuum (SC) output (red). The vertical axis stands for the optical power per comb tooth. The vertical lines show optical frequencies used to obtain the $f_{CEO}$ signal and the frequency ratios of optical lattice clocks.

**Fig. 3.** Modified Allan deviation of the frequency spectral transfer noise of the comb $\sigma_{Comb}$. The blue (red) line shows the transfer using a single-port Er:fiber comb from 215 to 259 THz (282 THz), the black line shows the transfer from 215 to 259 THz using a multibranch NPR Er:fiber comb, and the green line shows the in-loop signal of $f_{CEO}$. The error bars stand for 1-$\sigma$ statistical uncertainty. The circles and triangles show different measurement runs to evaluate short-term and long-term instabilities measured with the basic measurement intervals of 0.02 and 0.2 s, respectively.

**Fig. 4.** (a) Fractional frequency noise PSD observed in the optical frequency transfer. The black line shows the stability limit derived from the SNR of the beat signals with the phase noise floor of -80 dBc/Hz. The green line shows the frequency instability of an optical cavity reported in Ref. 26. The purple line assumes the thermal noise limit for 40-cm-long cavities with crystalline-coated mirrors. (b) Instabilities for Yb/Sr and Hg/Sr ratio measurements, assuming experimental conditions of a 'clock laser' instability with $1 \times 10^{-16}$ at 1 s and $T_C = T_i + 1$ s.

(a)

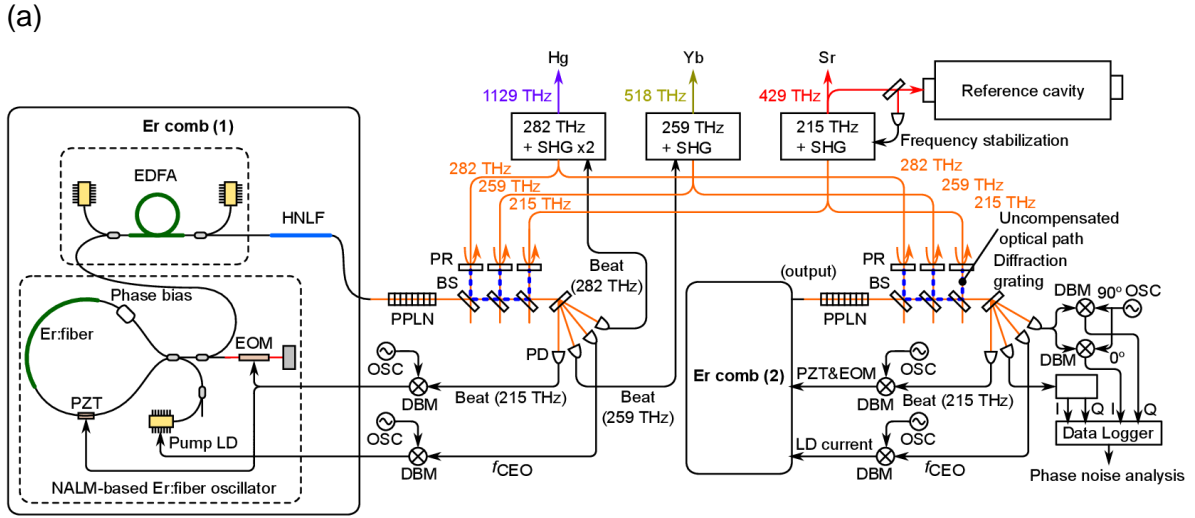

(b)

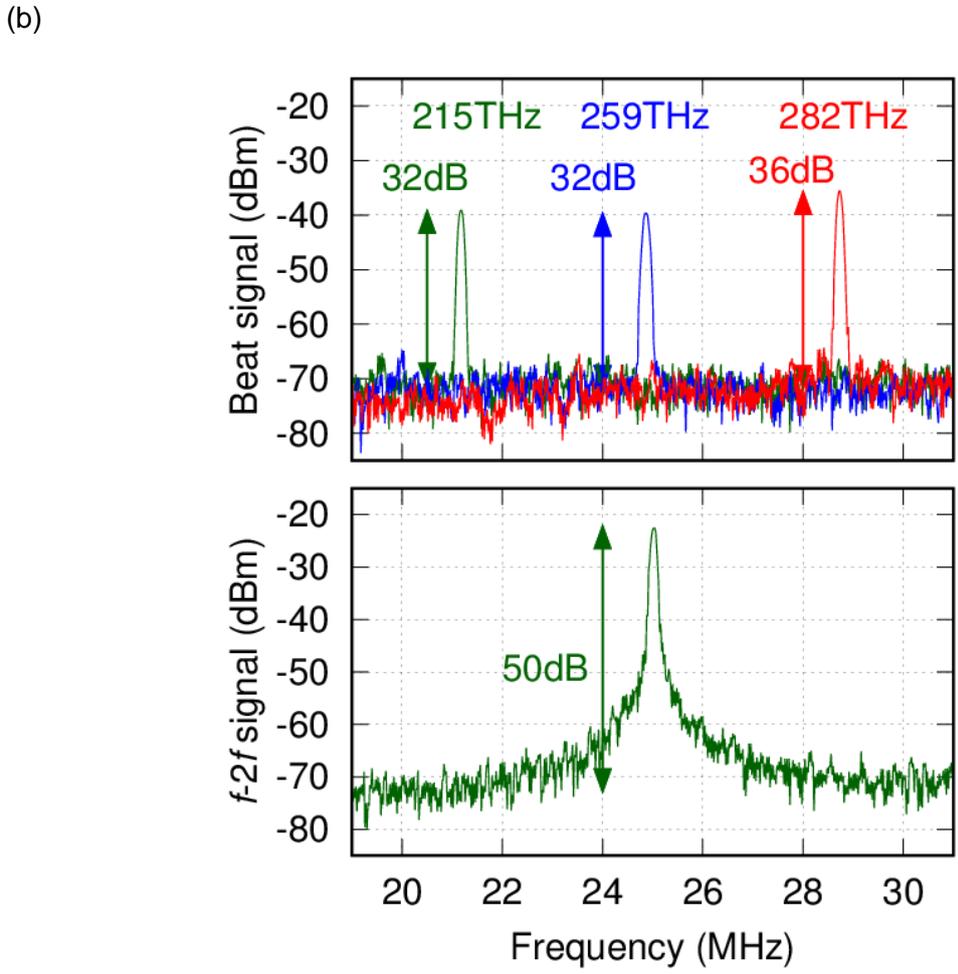

Fig.1.

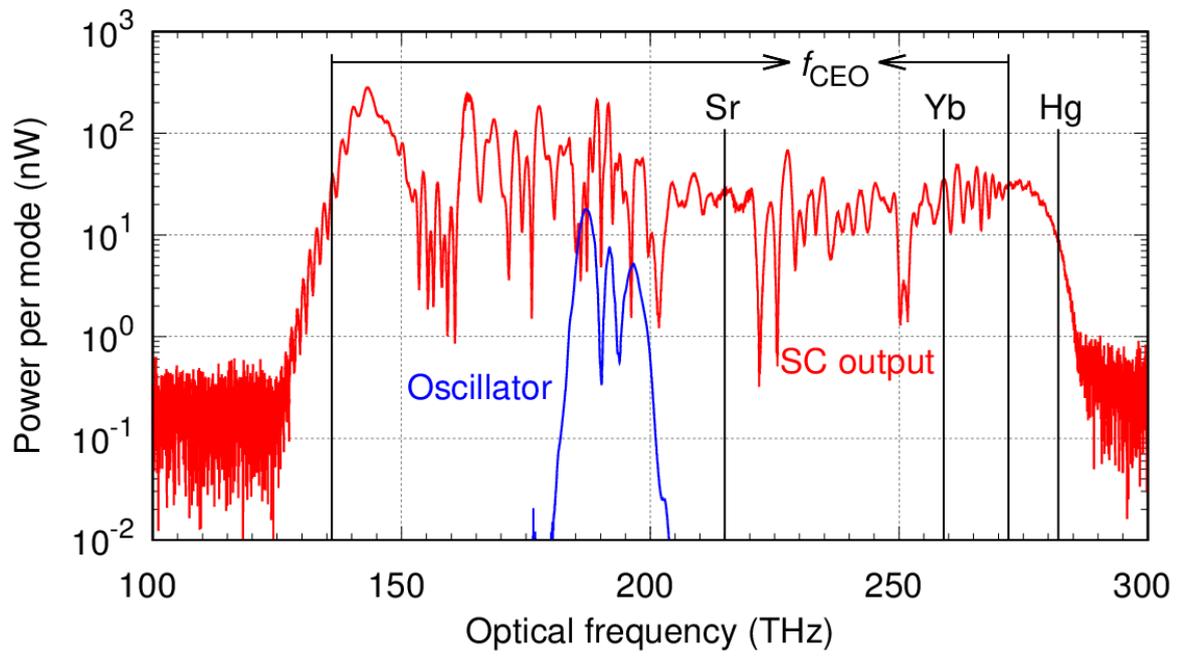

Fig. 2.

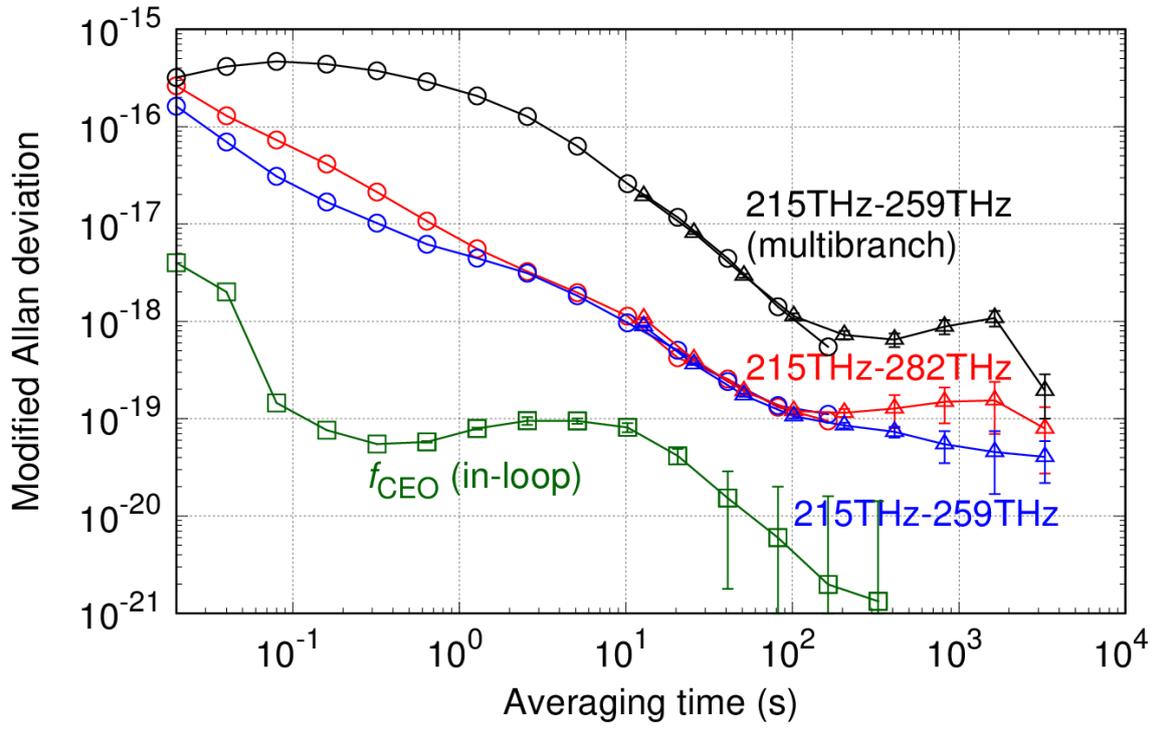

Fig. 3.

(a)

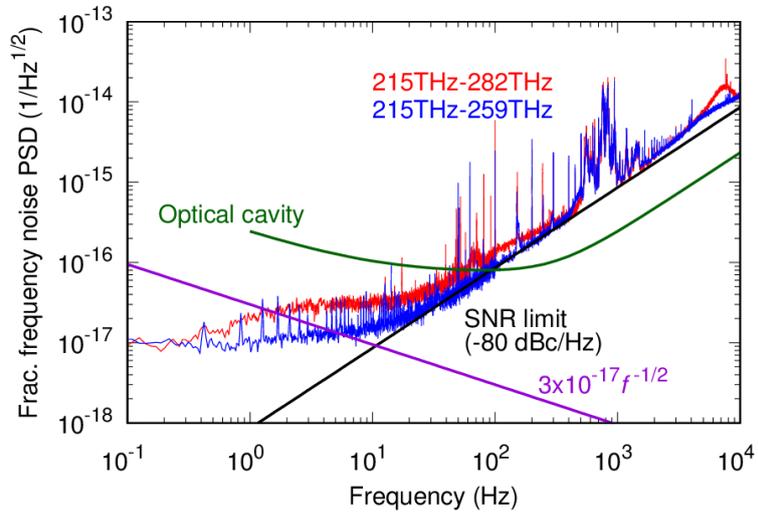

(b)

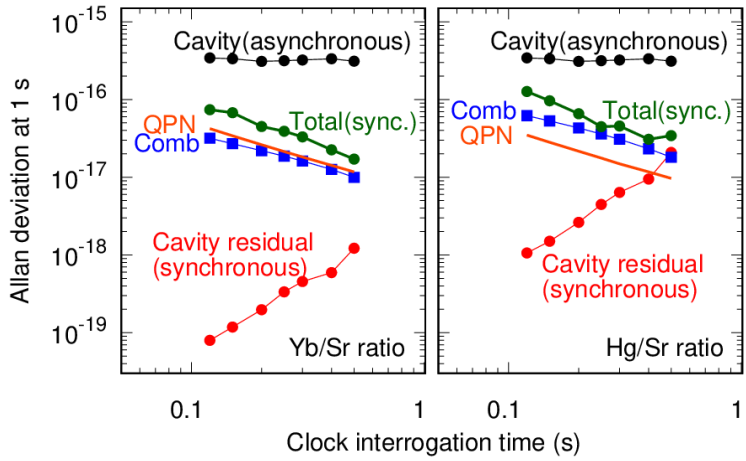

Fig. 4.